\documentclass{article}
\usepackage{spconf,amsmath,graphicx}
\usepackage{booktabs}
\usepackage{multirow}

\def\d{{\mathbf d}}
\def\h{{\mathbf h}}

\usepackage{amsmath}
\DeclareMathOperator*{\argmax}{arg\,max}

\title{Focus on the present: a regularization method for the ASR source-target attention layer}
\name{Nanxin Chen, Piotr \.{Z}elasko, Jes\'us Villalba, Najim Dehak}
\address{Center for Language and Speech Processing, Johns Hopkins University, Baltimore, MD, USA\\
\{nchen14, pzelask2, jvillal7, ndehak3\}@jhu.edu}
\begin{document}
\maketitle
\begin{abstract}
This paper introduces a novel method to diagnose the source-target attention in state-of-the-art end-to-end speech recognition models with joint connectionist temporal classification (CTC) and attention training.
Our method is based on the fact that both, CTC and source-target attention, are acting on the same encoder representations.
To understand the functionality of the attention, CTC is applied to compute the token posteriors given the attention outputs.
We found that the source-target attention heads are able to predict several tokens ahead of the current one.
Inspired by the observation, a new regularization method is proposed which leverages CTC to make source-target attention more focused on the frames corresponding to the output token being predicted by the decoder.
Experiments reveal stable improvements up to 7\% and 13\% relatively with the proposed regularization on TED-LIUM 2 and Librispeech. 
\end{abstract}
\section{Introduction}
\label{sec:intro}
In recent times, a lot of progress has been made to improve the end-to-end speech recognition systems~\cite{karita2019comparative, luscher2019rwth}.
End-to-end (E2E) systems have various advantages, such as simple training, joint optimization of different components, and scaling up easily to larger training data and resources.
However, there are substantial concerns about applying such models to real applications due to their black-box nature.
While the network can reach good overall performance, it still fails for certain examples.
Understanding how the network makes predictions and explaining why the network fails under certain cases are very important for this field.

In this paper, we are focusing on the key component of the end-to-end structure, the source-target attention.
Source target attention plays a major role in the E2E ASR and it has been widely used in LSTM-based E2E systems~\cite{chorowski2015attention,chan2016listen,watanabe2018espnet} and Transformer-based E2E systems~\cite{dong2018speech,karita2019comparative,luscher2019rwth}.
Most recent efforts on visualizing and understanding attention models have focused on calculating the contribution of the input unit to the final decision~\cite{Simonyan14a,mahendran2015understanding,nguyen2015deep,girshick2014rich,bach2015pixel,li2016visualizing}.
A lot of effort has been done to understand the attention mechanism for neural machine translation~\cite{bahdanau2015neural,ding2017visualizing,voita2018context,tang2018analysis}, and unsupervised pre-training~\cite{vig2019analyzing}.
However, for speech recognition, the input signal (i.e. acoustic frames) is different from the output sequence (i.e. text).
While visualization is still possible, the importance of individual frames is more difficult to interpret.
The existing studies on speech recognition try to extract information from hidden representations using an external classifier~\cite{belinkov2019analyzing}, unsupervised visualization~\cite{mohamed2012understanding}, or reconstruction~\cite{li2020does}.
Those methods are usually not architecture specific and relying on frame-level annotations.

In contrast, our analysis is based on state-of-the-art encoder-decoder speech recognition systems with joint connectionist temporal classification (CTC) and attention training~\cite{hori2017joint}.
Our method doesn't require training any external models and any pre-trained checkpoints can be used for visualization.
We used the trained CTC module to inspect the outputs from the source-target attention.
To understand what information the attention is providing, we compute the token posteriors by feeding the attention output embeddings to the CTC classifier.
Source-target attention heads return a weighted sum of encoder outputs and the CTC classifier is also trained on the same representations.
Thus, it is natural to use CTC classifier directly on the attention outputs, without the need to train an external model for inspection.
This provides us the ability to classify the attention outputs and understand how it contributes to the final decision making.

To the best of our knowledge, this is the first study to analyze the behavior of source-target attention inside the state-of-the-art Transformer models for speech recognition.
One typical assumption of the end-to-end model is that the model can utilize bi-directional information to make decisions.
In this study, we show that a well-trained E2E model can actually attend to frames corresponding to the token which is several tokens ahead.
The ability to inspect the output from attention inspired us to propose a novel regularization term assisting attention to build the proper mapping between frames and tokens.
While the overall mapping should be monotonic, it is not intuitive how different attention heads should behave.
Our regularization encourages some attention heads to focus on the frames corresponding to the current prediction.
Experiment results are conducted on several corpora and we observe stable improvements by adding this regularization without increasing the number of parameters.

\section{Joint CTC \& Attention training}

\begin{figure}[t]
    \centering
    \includegraphics[width=0.8\columnwidth]{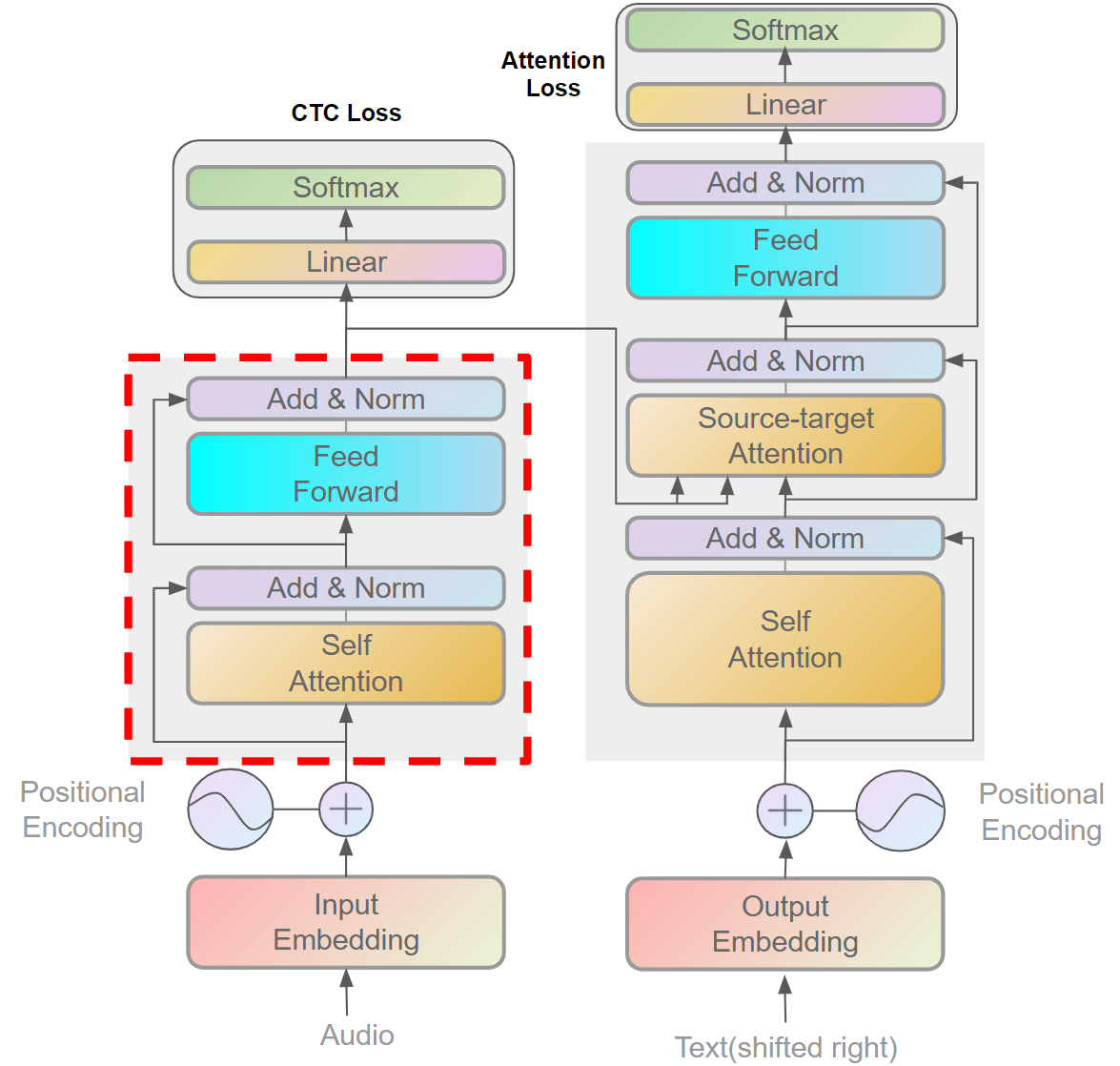}
     \vspace{-4mm}
    \caption{Diagram for joint CTC and attention training. The encoder outputs, highlighted by the red box, provides the input representations for both CTC and source-target attentions.}
    \label{fig:transformer}
   \vspace{-2mm}
\end{figure}

A popular framework used for End-to-end automatic speech recognition research is based on encoder-decoder architecture with joint connectionist temporal classification (CTC) and attention training~\cite{hori2017joint}, which is shown in Figure~\ref{fig:transformer}.
Here, we use the transformer network as an example but the same method can be also applied to other models, such as Long short-term memory (LSTM) network~\cite{chorowski2015attention,chan2016listen,watanabe2018espnet}.
The encoder extracts high-level audio-based representations from the input signal while the decoder combines information from both input signal and text context.
The interaction between encoder and decoder is achieved by the source-target attention located in the middle of the decoder transformer block.
More specifically, assume the output representation from encoder is $\h_1, \h_2, ..., \h_T$, the source-target attention can be considered as a weighted sum of the encoder outputs
\begin{equation}
    \mathbf{f}^{\mathrm{att}}_i = \sum_t w_{i,t} V(\h_t)
    \label{eq:att}
\end{equation}
where $i$ is the index of the output token and $V$ is a linear function known as value function.
The weight coefficient $w_{i,t}$ depends on both decoder state $s_i$ and encoder output $h_t$.
For the transformer based system, we have a $\mathbf{f}^{\mathrm{att}}_i$ per layer and head--we drop the indexes to keep the notation uncluttered.

In \cite{hori2017joint}, CTC is introduced to reduce the number of insertion and deletion errors from the attention prediction.
Pure attention-based ASR has been reported to be prone to include those errors because of its flexible alignment property, which can attend to any portion of the encoder outputs to predict the next label.
CTC probabilities are used to enforce a monotonic alignment that does not allow large jumps or looping of the same frames, which reduces insertion and deletion errors.

As shown in Figure~\ref{fig:transformer}, the CTC network includes one linear layer and softmax activation.
It also conditions on the encoder outputs, highlighted by the red box in the figure.

\vspace{-2mm}
\section{Understanding source-target attention by applying CTC classification}

Even though CTC enforces a monotonic alignment between the input frames and the output sequence, it still depends on the source-target attention to select the output tokens properly.
In other words, the source-target attention may fail to select some of the output tokens from encoder outputs.
In this work, we are focusing on the analysis of the source-target attention because of two important reasons:
\begin{itemize}
    \item Source-target attention is the only interaction between the decoder and input signal. 
    \item The behavior of the source-target attention is expected to be monotonic which is possible to diagnose, whereas no such expectation exists for self-attention.
\end{itemize}

\subsection{Using CTC posterior to analyze attention outputs}

In this paper, we are interested in the following embedding
\begin{equation}
    \d_i = \sum_t w_{i,t} \h_t
\end{equation}
Compare to the equation~\ref{eq:att}, $V$ function is ignored because it is both linear and data-independent.
Also by the design of the attention mechanism, all the weights should sum up to one %
\begin{equation}
    \sum_t w_{i, t} = 1
\end{equation}
So $\d_i$ is a weighted sum of $\h$.
For example, if $w_i$ is one-hot vector, $\d_i$ basically selects the corresponding $\h_t$.

To understand attention weights $w_{i,t}$, the alignment between $\h_t$ and decoder state $s_i$ is usually needed.
This is both time-consuming to get and difficult to analyze.
Instead, we focus on the embedding $\d_i$ since it is not position-related.
Since $\d_i$ is a weighted sum of $\h$, and the CTC is also trained on $\h$, our idea is to use CTC to analyze $\d_i$ for each attention head and each output token.
In our case, CTC includes only one linear layer and the posterior logits are given by
\begin{equation}
    l_{i,c} = \mathbf{W}^{\mathrm{CTC}}_c \d_i + \mathbf{b}^{\mathrm{CTC}}_c
    \label{eq:logits}
\end{equation}
for a given attention, where $c$ is the output token, $\mathbf{W}^{\mathrm{CTC}}$ and $\mathbf{b}^{\mathrm{CTC}}$ are the parameters for the CTC layer.

And the token posterior can be computed by
\begin{equation}
    p_{i,c} = \frac{\mathrm{e}^{l_{i,c}}}{\sum_{c^{'}} \mathrm{e}^{l_{i,c^{'}}}}
\end{equation}

Our analysis is concentrating on $\argmax_c p_{i,c}$ which provides us the token attention head found at step $i$.
This token can either be the real token included in the vocabulary or the special token $\epsilon$ defined for CTC objective.

\begin{table*}
\centering
\caption{Example of network predictions from E2E ASR without external language model.
The decoder includes six layers and each layer includes eight attention heads.
The blank cell indicates that the output from the head has been classified as $\epsilon$.
\_ indicates the start of the new word and
H$i$ is the $i$-th attention head.
Utterance ID 533-131564-0015: \it{I know they are bless them}}
\label{tab:example}
\scalebox{0.9}{
\begin{tabular}{@{}cccccccccccccccccc@{}}
\toprule
\multirow{2}{*}{\begin{tabular}[c]{@{}c@{}}Ground\\ Truth\end{tabular}} & \multirow{2}{*}{\begin{tabular}[c]{@{}c@{}}Network\\ Prediction\end{tabular}} & \multicolumn{8}{c}{Layer 5}                      & \multicolumn{8}{c}{Layer 6}                    \\ \cmidrule(l){3-18} 
                                                                        &                                                                               & H1 & H2  & H3     & H4 & H5    & H6    & H7 & H8 & H1      & H2 & H3 & H4    & H5 & H6  & H7 & H8 \\ \midrule
\_i                                                                     & \_i                                                                           &    & {\bf \_i} & {\sl \_know} &    & {\sl \_are} & \underline{{\sl ed}}    &    &    &         &    &    & {\sl \_are} &    & \underline{{\sl ed}}  &    &    \\
\_know                                                                  & \_know                                                                        &    &     &        &    &       &       &    &    &         &    &    &       &    &     &    &    \\
\_they                                                                  & \_they                                                                        &    &     & {\sl \_are}  &    &       &       &    &    &         &    &    &       &    & \_i &    &    \\
\_are                                                                   & \_are                                                                         &    &     &        &    &       & {\bf \_are} &    &    &         &    &    &       &    &     &    &    \\
\_bless                                                                 & \_bless                                                                       &    &     &        &    &       &       &    &    & {\bf \_bless} &    &    &       &    &     &    &    \\
\_them                                                                  & ed                                                                          &    &     &        &    &       &       &    &    &         &    &    &       &    &     &    &    \\
\textless{}eos\textgreater{}                                            & \textless{}eos\textgreater{}                                                  &    &     &        &    &       &       &    &    &         &    &    &       &    &     &    &    \\ \bottomrule
\end{tabular}}
\end{table*}

\vspace{-2mm}
\subsection{Example}

A decoding example is shown in Table~\ref{tab:example}.
We used the state-of-the-art transformer model~\cite{karita2019comparative} for visualization. The model is trained on Librispeech~\cite{panayotov2015librispeech} without any changes.
The model checkpoint is publicly accessible \footnote{https://drive.google.com/open?id=\\ 17cOOSHHMKI82e1MXj4r2ig8gpGCRmG2p}.

Considering the current token to predict, three different situations can be observed from this example:

\noindent\textbf{Predict forward} Italicized tokens are predicted ahead of their corresponding output time step.
For example, the network is able to predict the word {\it know}, {\it are}, {\it ed} without any text context during the prediction of the first token {\it i}.

\noindent\textbf{Predict present} Tokens predicted at their actual time step by the source-target attention are highlighted in bold.
When tokens are not detected at their corresponding time step by the source-target attention, the 
self-attention blocks are able to use the {\it predict forward} predictions to detect the present token.
For instance, ``know" was predicted at the beginning and the self-attention may select it 
for the second prediction.

\noindent\textbf{Predict backward} The head points to token from a past time step.
This case is redundant because, in auto-regressive decoding, past tokens are also provided as the decoder input.

Besides these three cases, a large number of attention outputs are classified as $\epsilon$, the special token in CTC.
Our assumption is that those outputs are not contributing to the final decisions since it is well-known that the large modern neural networks have a significant amount of redundancy~\cite{han2015learning}.

\subsection{Observations}

To better understand the role of different decoder layers, we carefully examine the number of unique tokens each layer found within each utterance.
This statistic is related to the length of the utterance but comparison among different layers is still meaningful.
The result is reported in Table~\ref{tb:obs}.

\begin{table}[t]
    \centering
    \caption{Number of unique tokens (including $\epsilon$) each layer found on average within one utterance.}
    \label{tb:obs}
    \scalebox{0.90}{
    \begin{tabular}{cc}
        \toprule
        \textbf{Decoder Layer Index} & \textbf{mean $\pm$ std} \\
        \midrule
        0 & $1.9 \pm 0.3$\\
        1 & $1.0 \pm 0.2$\\
        2 & $1.1 \pm 0.4$\\
        3 & $2.8 \pm 1.9$\\
        4 & $11.1 \pm 6.3$\\
        5 & $11.6 \pm 7.4$\\
        \bottomrule
    \end{tabular}}
    \vspace{-7pt}
\end{table}

From the table, source-target attention in the lower layers of the decoder can only find $1-2$ unique tokens on average within one utterance.
In comparison, the last two layers can find more than 10 different tokens.
This indicates the source-target attention in lower decoder layers is not strong and suggests that those layers are more responsible for language modeling.
Our observations are consistent with the previous studies~\cite{voita2018context,ramsauer2020hopfield}, where~\cite{voita2018context} tried to prune the number of attention heads and~\cite{ramsauer2020hopfield} inspects the shape of the attention weights.

\section{CTC-based regularization on source-target attention}

Based on the analysis in the last section, we propose a novel regularization which uses CTC classification results to force the source-target attention to select corresponding frames.

The source-target attention is encouraged to predict the current token but the question is which layer and which head should make the prediction.
For the deep transformer cases, this question is not answered yet in literature.
In this paper, we are trying to answer this question by allowing both the case of $\epsilon$ and "Predict present".
More specifically, we compute the overall focus of target token $y_i$ by maximizing $\mathrm{att}$ over all decoder layers and source-target attention heads
\begin{equation}
    \mathrm{focus}_{i,c} = \max_{\mathrm{att}} l^{\mathrm{att}}_{i,c} 
\end{equation}
where $l$ is defined in equation~\ref{eq:logits}.
The attention probability $\mathbf{q}_i$ normalizes over all tokens except $\epsilon$
\begin{equation}
    q_{i,c} = \frac{\mathrm{e}^{\mathrm{focus}_{i,c}}}{\sum_{c^{'} \neq \epsilon} \mathrm{e}^{\mathrm{focus}_{i,c^{'}}}}
\end{equation}
to allow the existence of $\epsilon$-predictions.

The regularization we proposed is the cross-entropy between the overall attention probability and target sequence
\begin{equation}
    L_{\mathrm{reg}} = - \lambda \sum_i \ln{q_{i,y_i}}
\end{equation}
Because the network can also predict ahead, we added a coefficient $\lambda$ to weight this regularization loss.
From our experiments results, this coefficient is important because when the regularization weight is high, the model is discouraged to predict tokens in the future and loses the ability to use the bi-directional contextual information.

In all experiments, $\mathbf{W}^{\mathrm{CTC}}$ and $\mathbf{b}^{\mathrm{CTC}}$ are updated only by the CTC loss.
We disconnect this regularization loss from the update of CTC parameters.

\begin{table}[t]
    \centering
    \caption{Word error rates (WERs) for Tedlium2.}
    \label{tb:tedlium2}
    \scalebox{0.90}{
    \begin{tabular}{lcc}
        \toprule
         \textbf{WER} & \textbf{dev} & \textbf{test} \\
        \midrule
        \quad Baseline & 9.3 & 8.1 \\
        \midrule
        {\sl Proposed} \\
        \quad $\lambda=0.1$ & {\bf 8.7} & 7.8 \\
        \quad $\lambda=0.2$ & 8.8 & {\bf 7.7}\\
        \quad $\lambda=0.3$ & 9.6 & 8.0 \\
        \bottomrule
    \end{tabular}}
    \vspace{-7pt}
\end{table}

\section{Experiments}

We conducted experiments on TED-LIUM 2~\cite{rousseau2014enhancing} and Librispeech~\cite{panayotov2015librispeech} using ESPnet~\cite{watanabe2018espnet,karita2019comparative}. 
We used 80 dimension mel-scale filterbank coefficients in addition to three-dimensional pitch features.
Following the standard setups, we applied speed perturbation, SpecAugment~\cite{park2019specaugment} for TED-LIUM 2 and SpecAugment for Librispeech.
For all experiments, we used Byte-Pair Encoding tokens~\cite{gage1994new} and the number of tokens is 500, 5000 for TED-LIUM 2 and Librispeech.

In all experiments, the encoder included 12 blocks starting with convolutional layers for down-sampling.
The decoder consisted of 6 blocks which included 4 attention heads, 256 hidden units and 2048 feed-forward units.
The network was trained for 100 epochs with \textbf{single} NVIDIA GTX1080TI GPU.
Warmup \cite{vaswani2017attention} is used for the first 25,000 steps.

TED-LIUM 2 results are given in Table~\ref{tb:tedlium2}.
Improvements are observed when the weight of the regularization is less than $0.2$.
When $\lambda > 0.2$, it actually hurts the performance.
Word error rate improved $6.5\%$ and $3.7\%$ relatively on development and test set, respectively, with the best configuration.

\begin{table}[t]
    \centering
    \caption{Word error rates (WERs) for Librispeech trained on a single NVIDIA 1080TI GPU.}
    \label{tb:librispeech}
    \begin{tabular}{lcccc}
        \toprule
         \textbf{WER} & \textbf{dev\_clean} & \textbf{dev\_other} & \textbf{test\_clean} & \textbf{test\_other} \\
        \midrule
        \quad Baseline & {\bf 3.7} & 9.8 & {\bf 4.0} & 10.0 \\
        \midrule
        {\sl Proposed} \\
        \quad $\lambda=0.1$ & 3.8 & {\bf 8.6} & 4.1 & {\bf 8.7} \\
        \bottomrule
    \end{tabular}
    \vspace{-7pt}
\end{table}

Librispeech result is given in Table~\ref{tb:librispeech} and we adopted the best configuration $\lambda=0.1$ from TED-LIUM 2 results.
We set the CTC weight to $0.05$ and language model (LM) weight to $0.4$ for decoding which is tuned on development set.
We found that smaller CTC and LM weight works better which suggests that predictions from the decoder becomes more important by incorporating this regularization.
Results on clean conditions are comparable with the baseline performance, while we observed large improvements on the ``noisy" other set.
We believe the proposed regularization helps attention to locate the token under noisy conditions.
Word error rate reduced by $12.2\%$ and $13.0\%$ respectively on development other and test other set.

\vspace{-3mm}
\section{Conclusion}
In this paper, we present a novel way to understand the functionality of the source-target attention by inspecting the weighted sum from the attention outputs.
Connectionist temporal classification is applied to analyze this weighted sum.
From our observations on the state-of-the-art transformer speech recognition model, the source-target attention not only can select frames of the current token but it also can consider frames of the future tokens.
This provides a new direction to understand the black box model for end-to-end speech recognition research.
We further propose a novel CTC-based regularization to enforce the source-target attention to focus on corresponding frames.
Our regularization can be applied to state-of-the-art transformer models and it brings improvements on several datasets.
Future research includes understanding $\epsilon$-predictions, and adapting the regularization to support multiple cases.

\bibliographystyle{IEEEbib-abbrev}
\bibliography{strings,refs}

\begin{thebibliography}{10}

\bibitem{karita2019comparative}
S.Karita, N.Chen, T.Hayashi, T.Hori, H.Inaguma, Z.Jiang, M.Someki,
  N.~E.~Y.Soplin, R.Yamamoto, X.Wang, et~al.,
\newblock ``A comparative study on transformer vs rnn in speech applications,''
\newblock {\em Proc. ASRU}, pp. 449--456, 2019.

\bibitem{luscher2019rwth}
C.L{\"u}scher, E.Beck, K.Irie, M.Kitza, W.Michel, A.Zeyer, R.Schl{\"u}ter, and
  H.Ney,
\newblock ``Rwth asr systems for librispeech: Hybrid vs attention,''
\newblock {\em Proc. Interspeech}, pp. 231--235, 2019.

\bibitem{chorowski2015attention}
J.~K.Chorowski, D.Bahdanau, D.Serdyuk, K.Cho, and Y.Bengio,
\newblock ``Attention-based models for speech recognition,''
\newblock in {\em Advances in neural information processing systems}, 2015, pp.
  577--585.

\bibitem{chan2016listen}
W.Chan, N.Jaitly, Q.Le, and O.Vinyals,
\newblock ``Listen, attend and spell: A neural network for large vocabulary
  conversational speech recognition,''
\newblock {\em Proc. ICASSP}, pp. 4960--4964, 2016.

\bibitem{watanabe2018espnet}
S.Watanabe, T.Hori, S.Karita, T.Hayashi, J.Nishitoba, Y.Unno, N.-E.~Y.Soplin,
  J.Heymann, M.Wiesner, N.Chen, et~al.,
\newblock ``Espnet: End-to-end speech processing toolkit,''
\newblock {\em Proc. Interspeech}, pp. 2207--2211, 2018.

\bibitem{dong2018speech}
L.Dong, S.Xu, and B.Xu,
\newblock ``Speech-transformer: a no-recurrence sequence-to-sequence model for
  speech recognition,''
\newblock {\em Proc. ICASSP}, pp. 5884--5888, 2018.

\bibitem{Simonyan14a}
K.Simonyan, A.Vedaldi, and A.Zisserman,
\newblock ``Deep inside convolutional networks: Visualising image
  classification models and saliency maps,''
\newblock in {\em Workshop at ICLR}, 2014.

\bibitem{mahendran2015understanding}
A.Mahendran and A.Vedaldi,
\newblock ``Understanding deep image representations by inverting them,''
\newblock in {\em Proceedings of CVPR}, 2015, pp. 5188--5196.

\bibitem{nguyen2015deep}
A.Nguyen, J.Yosinski, and J.Clune,
\newblock ``Deep neural networks are easily fooled: High confidence predictions
  for unrecognizable images,''
\newblock in {\em Proceedings of CVPR}, 2015, pp. 427--436.

\bibitem{girshick2014rich}
R.Girshick, J.Donahue, T.Darrell, and J.Malik,
\newblock ``Rich feature hierarchies for accurate object detection and semantic
  segmentation,''
\newblock in {\em Proceedings of CVPR}, 2014, pp. 580--587.

\bibitem{bach2015pixel}
S.Bach, A.Binder, G.Montavon, F.Klauschen, K.-R.M{\"u}ller, and W.Samek,
\newblock ``On pixel-wise explanations for non-linear classifier decisions by
  layer-wise relevance propagation,''
\newblock {\em PloS one}, vol. 10, no. 7, pp. e0130140, 2015.

\bibitem{li2016visualizing}
J.Li, X.Chen, E.Hovy, and D.Jurafsky,
\newblock ``Visualizing and understanding neural models in nlp,''
\newblock in {\em Proceedings of NAACL-HLT}, 2016, pp. 681--691.

\bibitem{bahdanau2015neural}
D.Bahdanau, K.Cho, and Y.Bengio,
\newblock ``Neural machine translation by jointly learning to align and
  translate,''
\newblock in {\em Proceedings of ICLR}, 2015.

\bibitem{ding2017visualizing}
Y.Ding, Y.Liu, H.Luan, and M.Sun,
\newblock ``Visualizing and understanding neural machine translation,''
\newblock in {\em Proceedings of ACL}, 2017, pp. 1150--1159.

\bibitem{voita2018context}
E.Voita, P.Serdyukov, R.Sennrich, and I.Titov,
\newblock ``Context-aware neural machine translation learns anaphora
  resolution,''
\newblock in {\em Proceedings of ACL}, 2018, pp. 1264--1274.

\bibitem{tang2018analysis}
G.Tang, R.Sennrich, and J.Nivre,
\newblock ``An analysis of attention mechanisms: The case of word sense
  disambiguation in neural machine translation,''
\newblock in {\em Proceedings of the Third Conference on Machine Translation:
  Research Papers}, 2018, pp. 26--35.

\bibitem{vig2019analyzing}
J.Vig and Y.Belinkov,
\newblock ``Analyzing the structure of attention in a transformer language
  model,''
\newblock {\em arXiv preprint arXiv:1906.04284}, 2019.

\bibitem{belinkov2019analyzing}
Y.Belinkov, A.Ali, and J.Glass,
\newblock ``Analyzing phonetic and graphemic representations in end-to-end
  automatic speech recognition,''
\newblock {\em Proc. Interspeech}, pp. 81--85, 2019.

\bibitem{mohamed2012understanding}
A.-r.Mohamed, G.Hinton, and G.Penn,
\newblock ``Understanding how deep belief networks perform acoustic
  modelling,''
\newblock {\em Proc. ICASSP}, pp. 4273--4276, 2012.

\bibitem{li2020does}
C.-Y.Li, P.-C.Yuan, and H.-Y.Lee,
\newblock ``What does a network layer hear? analyzing hidden representations of
  end-to-end asr through speech synthesis,''
\newblock {\em Proc. ICASSP}, pp. 6434--6438, 2020.

\bibitem{hori2017joint}
T.Hori, S.Watanabe, and J.~R.Hershey,
\newblock ``Joint ctc/attention decoding for end-to-end speech recognition,''
\newblock in {\em Proceedings of ACL}, 2017, pp. 518--529.

\bibitem{panayotov2015librispeech}
V.Panayotov, G.Chen, D.Povey, and S.Khudanpur,
\newblock ``Librispeech: an asr corpus based on public domain audio books,''
\newblock {\em Proc. ICASSP}, pp. 5206--5210, 2015.

\bibitem{han2015learning}
S.Han, J.Pool, J.Tran, and W.Dally,
\newblock ``Learning both weights and connections for efficient neural
  network,''
\newblock in {\em Advances in neural information processing systems}, 2015, pp.
  1135--1143.

\bibitem{ramsauer2020hopfield}
H.Ramsauer, B.Schäfl, J.Lehner, P.Seidl, M.Widrich, L.Gruber, M.Holzleitner,
  M.Pavlović, G.~K.Sandve, V.Greiff, D.Kreil, M.Kopp, G.Klambauer,
  J.Brandstetter, and S.Hochreiter,
\newblock ``Hopfield networks is all you need,''
\newblock {\em arXiv preprint arXiv:2008.02217}, 2020.

\bibitem{rousseau2014enhancing}
A.Rousseau, P.Del{\'e}glise, Y.Esteve, et~al.,
\newblock ``Enhancing the ted-lium corpus with selected data for language
  modeling and more ted talks.,''
\newblock in {\em LREC}, 2014, pp. 3935--3939.

\bibitem{park2019specaugment}
D.~S.Park, W.Chan, Y.Zhang, C.-C.Chiu, B.Zoph, E.~D.Cubuk, and Q.~V.Le,
\newblock ``{SpecAugment: A Simple Data Augmentation Method for Automatic
  Speech Recognition},''
\newblock {\em Proc. Interspeech}, pp. 2613--2617, 2019.

\bibitem{gage1994new}
P.Gage,
\newblock ``A new algorithm for data compression,''
\newblock {\em C Users Journal}, vol. 12, no. 2, pp. 23--38, 1994.

\bibitem{vaswani2017attention}
A.Vaswani, N.Shazeer, N.Parmar, J.Uszkoreit, L.Jones, A.~N.Gomez, {\L}.Kaiser,
  and I.Polosukhin,
\newblock ``Attention is all you need,''
\newblock in {\em Advances in neural information processing systems}, 2017, pp.
  5998--6008.

\end{thebibliography}

\end{document}